\documentstyle[aps,preprint]{revtex}
\input epsf
\global\firstfigfalse  

\tighten

\def\NN{{\cal N}}

\def\none{$\NN=1$}
\def\noneplus{$\NN=1^*$}
\def\ntwo{$\NN=2$}
\def\nfour{$\NN=4$}

\def\tr{{\rm tr}}
\def\half{{1\over2}}

\def\tF{{\widetilde F}}
\def\a{{\alpha}}

\def\s{{\sigma}}
\def\t{{\tau}}
\def\de{{\delta}}
\def\Gp{{G_\parallel}}

\def\bel{\begin{equation}\label}
\def\ee{\end{equation}}
\def\Im{{\rm Im}}
\newcommand\eref[1]{(\ref{#1})}

\renewcommand{\*}{ &=& }

\newcommand{\OL}[1]{ \hspace{1pt}\overline{\hspace{-1pt}#1
   \hspace{-1pt}}\hspace{1pt} }

\begin{document}

\preprint{
\begin{minipage}[t]{3in}
\begin{flushright} MIT-CTP-3003
\\
hep-th/0007250
\\
July 2000
\end{flushright}
\end{minipage}
}

\title{
Finite Temperature Effects in  the Supergravity Dual of the $N=1^*$ 
Gauge Theory}

\author{Daniel Z. Freedman}
\address{Department of Mathematics and 
Center for Theoretical Physics\\
Massachusetts Institute of 
Technology, Cambridge, MA 02139} 
\author{Joseph A. Minahan\footnote{Address after Sept. 1, 2000: Department of 
Theoretical Physics, Uppsala University, Uppsala, Sweden}}
\address{
Center for Theoretical Physics \\
Massachusetts Institute of 
Technology, Cambridge, MA 02139}

\maketitle

\begin{abstract}
We consider the supergravity dual of the \noneplus\ theory at finite 
temperature by applying the Polchinski-Strassler construction to
the black D3 brane solution of Type IIB supergravity. At finite
temperature the 5-brane probe action is minimized when the probe falls to
the horizon, although metastable minima with $r>>r_H$ persist for a range
of temperatures. Thermal effects on the 3-form source for the hypermultiplet 
mass $m$ and its order $m^2$ back reaction on the other fields of the IIB 
theory are computed.  We find unique solutions which are regular at the 
horizon and have the correct behavior on the boundary. For fixed temperature 
$T$, the horizon shrinks for increasing $m^2$ suggesting that there is a 
critical temperature separating the system into high and low temperature 
phases. In the high temperature phase 5-branes are unnecessary since there 
are no naked singularities.   Using the order $m^2$  correction to the horizon 
area we calculate the correction to the entropy  to be 
$\Delta {\cal S}=-0.1714N^2m^2T$, 
which is less than the free field result.

\end{abstract}

\newpage
\pagestyle{plain}
\narrowtext
\baselineskip=18pt

\setcounter{footnote}{0}

\section{Introduction}

The Maldacena conjecture \cite{Maldacena:1998re} relating classical supergravity to strongly coupled 
conformal gauge theories has had many successes \cite{Aharony:2000ti}.  
Perhaps more remarkable
is the fact that one can apply it to nonconformal theories as well.  The first,
and perhaps most famous application has been to \nfour\ super Yang-Mills
at finite temperature \cite{Witten:1998zw}.  
One may consider this as a gauge theory compactified
on a circle where the fermions have  anti-periodic boundary conditions,
breaking the supersymmetry.  This dimensionally reduced gauge theory is
expected to be confining with a mass gap.  By arguing that this theory has
a supergravity dual which is a black hole in a curved space,  Witten
was able to demonstrate both the existence of an area law and a mass gap
in the strong coupling limit.

One would like to find other models that exhibit confining behavior.  
Such examples have been found, either by adding spin to the black
hole,
 breaking some of the global symmetries of the gauge theories
\cite{Russo:1999mm,Csaki:1999cb,Russo:1999by,Cvetic:1999ne,Cvetic:1999rb}, 
or by
considering more exotic theories such as Type 0 string theories 
\cite{Klebanov:1998yy,Minahan:1999tm,Klebanov:1999yy,Minahan:1999yr} or
even type II theories with a dilaton turned on
\cite{Kehagias:1999tr,Gubser:1999pk,Kehagias:1999iy}, breaking the \nfour\ 
supersymmetry.  In these latter cases, while there appear to be gaps
and area laws, there are also naked singularities which would seem to
destroy the viability of these theories.
However, singularities are not necessarily disasters.  Some of them may be 
hidden behind horizons \cite{Gubser:2000nd}.  Alternatively, they may be
resolved by stringy considerations.

Recently, Polchinski and Strassler considered another model  \cite{ps}, 
the \noneplus\
model, in which mass terms for the three chiral multiplets that make up an
\nfour\ multiplet are included.  Several interesting generalizations 
of \cite{ps} have also appeared
\cite{Bena:2000zb,Bena:2000fz,Bena:2000va,Bachas:2000dx,Zamora:2000,Frey:2000be}. 
At high energies, the theory
is the usual \nfour\ model, while at low energies, the supersymmetry is 
broken to \none.  Unlike gauge theories with higher supersymmetries,
\none\ theories can be confining.  In fact, this gauge theory has a
large class of degenerate but discrete vacua \cite{Vafa:1994tf,Donagi:1996cf}
which include
Higgs, confining, and oblique confining phases. In these cases there is a
mass gap, but there are also vacua with unbroken $U(1)$ subgroups 
and thus massless photons.

Polchinski and Strassler constructed the supergravity dual of this model
as follows.  They assumed that a 3-form field strength is turned on
whose strength is proportional to the mass $m$ of the hypermultiplets.  The
field strength induces a Myers dielectric effect on the D3 branes, essentially
expanding them out in the transverse directions
to a two sphere\cite{Myers:1999ps,Kabat:1998im}.  
The two sphere is effectively a D5 
brane,
or one of its $S$-duals, with $n$ units of magnetic flux.  These flux units
minimally couple to the 4-form Ramond-Ramond gauge potential, and hence they 
correspond
to D3 brane charges.  The background metric, for large distances, that is for
the UV limit, is dominated by the D3 brane charges.  But as one moves toward
the IR, one starts probing near the D5 brane and here, the metric is dominated
by the D5 brane.  In between the metric interpolates between these two limits.

The presence of this D5 brane significantly changes the physics \cite{ps}.  
For example,
at large separations, magnetic charges are effectively confined, since a 
D string will have finite tension as it approaches the D5 brane.  However,
the electric charges are screened, since the fundamental string can end on
the D5 brane.  Likewise, for the S-dual picture, electric charges are confined
since a fundamental string has finite tension near an NS5-brane, but the
magnetic charges are screened because the D-string can end on the NS5-brane.

Polchinski and Strassler also demonstrated how the supergravity duals map into
the various degenerate vacua of the \noneplus\ theory.
The D3 branes can expand into more than one 5 brane and in the regime where
the supergravity limit {\it is valid}, 
one can find a one to one correspondence 
between the various ways of dividing up D3 branes into 5 branes and the 
degenerate vacua. 
For these vacua, one expects a gap, in 
other words, the absence of massless particles.
Actually, this is not completely true, since many vacua 
have unbroken U(1) gauge 
groups.  However, the number of massless states is at most of order 
$\sqrt{N}$ and their effects are not seen in supergravity duals.

If one were to consider the \noneplus\ theory at finite temperature, 
then the multiple vacua and gaps have profound effects. Consider first the
entropy. The entropy of
a massless theory in four dimensions scales
as $T^3$.  However, a theory with a gap will have its entropy suppressed
exponentially when the temperature falls below the mass gap scale. 
In the large $N$ limit,  
this should occur as a phase transition between
high and low temperature phases, where at some critical
temperature the free energies of the two phases is equalized. 
Another effect of finite temperature is that the degeneracy between the 
separate vacua should be lifted, leading to many metastable states.  
As the temperature is raised above the mass gap scale,  these
metastable vacua should disappear altogether, and at very high temperatures,
the theory should behave just like the \nfour\ theory at finite temperature.

In the supergravity dual of these theories, the high temperature phase
is described by a black hole in an asymptotic $AdS$ space. At high temperatures
there is a unique vacuum.  The multiple vacuum configurations of 
5-branes
described by Polchinski and Strassler will now have a higher free energy than
the vacuum.  The 5-branes were introduced
to account for naked singularities that inevitably appear and which were
specifically discussed in  \cite{gppz}. 
However,  
at high temperatures, singularities are shielded by the horizon; hence,
there is no need for the 5-branes.  

For the high temperature phase,
if one considers a spherical 5 brane probe with some D3 brane charge, then
 one should find that the free energy is minimized when the 5 brane
lies on the horizon.  One should also find that the energy is the same as
if there were only D3 branes and no 5 branes, since the 5 brane has no
net charge, and hence would be undetectable at the horizon.

However, in the high temperature phase, there still may exist local minima
for the 5-branes outside the horizon.  These configurations correspond to
metastable vacua, valid in the probe limit described in \cite{ps},
and which can be in a partial Higgs, confining or
oblique confining phase.  As the temperature is increased, these local minima 
eventually disappear.  

A crucial step for going beyond the probe calculation is the realization
that the probe calculation itself is unchanged for an arbitrary distribution of
parallel D3 branes \cite{ps}.  
Hence, one can construct a spherical
shell of D3 branes, of
 order $N$ in number, and still use the probe calculation to
determine the radius of the shell.  
This is not to say that the exact supergravity solution has been constructed,
since one still needs to compute the back reaction.  But it does allow one
to build an interpolating metric between a D3 brane geometry and a 
5-brane geometry.  However, in the finite temperature case, one cannot 
use this approach as a starting point,
since a configuration of D3 branes outside the horizon is not a solution
to Einstein's equations.

One can also explore the effects of the hypermultiplet mass on the entropy 
in the high temperature phase.
If the hypermultiplets are given a small mass $m$, then
there are polynomial corrections to the entropy.
In particular, to order $m^2$ the free field limit of the entropy is given by
\begin{equation}\label{entropy}
{\cal S}= \frac{2\pi^2}{3} N^2 T^3-\frac{3}{4}N^2m^2T + {\rm O}(m^4)
\end{equation}
The strong coupling calculation for the massless case
was carried out in \cite{gkp96}, where the area of a blackhole in AdS space 
was computed.  These authors found that
\begin{equation}\label{entropy0}
{\cal S}_0= \frac{\pi^2}{2} N^2 T^3,
\end{equation}  
a factor of $3/4$ from the free theory result.  An immediate question is to 
find the massive corrections  in the supergravity
limit and compare this with the free field result.

In this paper we study the finite temperature effects
outlined above 
for the
supergravity dual of the \noneplus\ theory. Our starting point is
 the black D3 brane solution
\cite{Horowitz:1991cd} of Type IIB supergravity perturbed by
 a 3-form gauge potential. A 5-brane probe
calculation  quickly reveals several features of finite temperature.
First, the $T=0$ minima \cite{ps} for  5-branes wrapping 
 2-spheres of  particular nonzero radii are now only  local minima; 
the true minimum appears at the horizon,
$r=r_H$ where the free energy is minimized. The local minima, which can be
thought of as metastable vacua, move to 
smaller radii and eventually disappear as the temperature is increased
, but at temperatures that can be much higher than when
the field theory is weakly coupled. We
demonstrate that at the horizon a wrapped 5-brane, which carries $n$
units of D3-brane charge, cannot be distinguished from a configuration of
$n$ D3 branes.

Although the major thermal effects in the probe calculation can be obtained
quite simply, other questions require the finite temperature modification
of the fields of the Type IIB theory through order $m^2$. We therefore 
compute exactly the 3-form field strength in the black D3 brane
background. The metric, 5-form field strength
 and dilaton are then sourced at order $m^2$
by bilinears in the 3-form, and we solve the equations which determine the
back reaction to this order.  Given the modified metric, the relation 
between temperature
$T$ and horizon distance $r_H$ is recomputed, and we find that the horizon
shrinks for fixed $T$ by an amount of order $\Delta r_H \sim m^2 R^2/T$ where
$R$ is the $AdS$ scale.   We presume that a more complete
calculation would show that $r_H \rightarrow 0$ at some 
temperature $T_c$ which is then the critical temperature for the phase
transition in the theory.

The computations for the back reaction can be specialized to the T=0 case,
where we must require that the homogeneous modes agree in their $SO(6)$
representation and $r$-dependence at the boundary with fields of the
dimensionally reduced $IIB$ supergravity \cite{krvn}. In general the
order $m^2$ inhomogeneous solutions approach the boundary at different
rates than the homogeneous modes. For example, the dilaton perturbation
equation is that of the second Kaluza-Klein excitation, the $\Delta=6$
field in the {\bf 20'} representation with an order $m^2$ source. The {\bf 20'}
representation agrees with the general relation between the $D=10$ and the
$D=5$ dilaton postulated in \cite{Pilch:2000ue}, but the solution does not
agree with the order $m^2$ term in the interpolating dilaton of
\cite{ps}.
The back reaction equations of metric components transverse to the
$D3$-branes describe scalars with $\Delta=8$ in the singlet
representation and $\Delta=6$ in the {\bf 84} representation of $SO(6)$.
We also show that the back reaction on the metric and 5-form lead precisely
to the order $m^2$ terms in the probe action, which were inserted in
\cite{ps} as a requirement of supersymmetry. All of these $T=0$ results
are a useful check that the rather complicated back reaction equations
are correct.

We also compute the decreased area of the horizon in the (approximate) 
modified geometry and thus obtain the entropy 
\begin{equation}\label{entrop}
{\cal S}_0= \frac{\pi^2}{2} N^2 T^3 -0.1714 N^2 m^2 T + {\rm O}(m^4)\
\end{equation}
including the order $m^2$ correction. Comparing with \eref{entropy} we see
that the large $N$ correction is 23\% 
of the free field result.

Technically speaking, it is noteworthy that finite temperature effects
can be computed essentially analytically as far as we have gone, and this
raises the hope of going beyond the leading order $m^2$ approximation.
At $T=0$ this proved to be easy \cite{ps} since the probe calculation is
not altered when the D3 branes are expanded around 2-spheres. This
fortuitous simplification is unlikely at finite $T$, since the only
solution to the  equations of motion when $m=0$ has all D3 branes at the
origin.

In Section II we consider the probe calculation.   In Section III
we compute the linear solutions for the 3-form field strengths in 
a finite temperature D3 background.  In Section IV  we obtain the back 
reaction  to the dilaton field. Order $m^2$ temperature dependent corrections
to the metric and 5-form are discussed in Section V, and the entropy
calculation is presented in Section VI.
The appendix contains a more detailed calculation
for the disappearance of metastable vacua.

\section{Probe Calculation for the Vacua}

In this section\footnote{the notation of \cite{ps} is used throughout the
present paper} we look for nontrivial solutions for a D5 brane probe
in a finite temperature D3 brane background geometry. 
Temperature is introduced via the thermal $D3$-brane
solution \cite{Horowitz:1991cd} of Type $IIB$ supergravity. The metric is
\begin{eqnarray}\label{metric}
ds^2\* (Z(r))^{-\frac{1}{2}}[-f(r)dt^2 + d{\bf x}^2] + 
       Z(r)^{\frac{1}{2}}[\frac{dr^2}{f(r)} + r^2 d\Omega^2_5]\\
Z(r)\* \frac{R^4}{r^4}\\              
f(r)\* 1-\frac{r_H^4}{r^4}\
\end{eqnarray} 
where $ R^4= 4 \pi g N \alpha'^2$. The solution
has a horizon at $r=r_H$, which corresponds to the temperature 
$T=r_H/(\pi R^2)$. The accompanying dilaton, axion and self-dual 5-form
are given by \cite{ps}
\begin{eqnarray}\label{dil}
e^\Phi \* g\\
\label{axion}
C\*\frac{\theta}{2\pi}\\
\label{f5}
\tilde{F}_{\it5} \*d\chi_{\it4} + *d\chi_{\it4}\\        
\chi_{\it4} \*\frac{dx^0\wedge dx^1\wedge dx^2\wedge dx^3}{gZ(r)}.\label{chieq}
\end{eqnarray}
While thermal effects cancel out for the $*$ operation contained in
 $\tilde{F}_{\it5}$, they will play an important role when $*$ acts on
other field strengths.

The action for the D5 brane is the sum of Born-Infeld and Chern-Simons
parts, and is given by   
\begin{equation}\label{D5action}
S = -\frac{\mu_5}{g}
\int d^6\xi\, \Bigl[-\det (G_\parallel)
\det(g^{-1/2}e^{\Phi/2}G_\perp + 2\pi\alpha'{\cal F})\Bigr]^{1/2}
+ \mu_5\int ( C_{\it 6} + 2\pi\alpha'{\cal F}_{\it 2} \wedge
C_{\it 4} )
\ , \label{d5act}
\end{equation}
where
\begin{equation}
2\pi\alpha'{\cal F}_{\it 2} = 2\pi\alpha'F_{\it 2} - B_{\it
2}\ .
\end{equation}
As in \cite{ps}, $G_\parallel$ refers to the pullback of the metric along
the ${ R^4}$, while $G_\perp$ is the pullback of the metric onto the
$S^2$ on which the 5-brane is wrapped. $C_{\it4}$ is the Ramond-Ramond gauge 
potential of the D3 branes, while
$C_{\it6}$ and $B_{\it2}$ are potentials for
the 3-form perturbation of the
background dual to the mass operator $m~tr(\Phi_1^2 +\Phi_2^2+\Phi_3^2)$ of 
the 
\noneplus\ field theory. 

The field $F_{\it2}$ is the $U(1)$ gauge field strength of the wrapped 5-brane,
and is assumed to contain $n$ units of flux, so that 
\begin{equation}\label{flux}
\int_{S^2}F_{\it2} = 2\pi n.
\end{equation}
Since $F_{\it2}$ couples to $C_{\it4}$ in the Chern-Simons term of \eref{d5act},
$n$ is the effective D3 charge of the 5-brane. 
As in \cite{ps}, we assume that the first term in \eref{d5act}\ is dominated
by $F_{\it2}$. The condition $n^2>>gN$ must then hold in order for the 
probe calculation to be valid \cite{ps}.

In the zero temperature limit, the leading order $n/\sqrt{gN}$ 
term from the Born-Infeld part
of \eref{d5act} cancels with the $C_{\it4}$ term in the Chern-Simons part.  
The radius of
the $S^2$ on which the D5 brane resides is then chosen to minimize  
the sum of first order correction terms.  
At finite temperature, leading terms no longer cancel 
because $\det G_\parallel$
is modified by the Schwarzschild factor \cite{Tseytlin:1999cq}, viz.
\begin{equation}\label{Gpar}
\det(G_\parallel) = Z^{-2}(1-r_H^4/r^4).
\end{equation}
There is then a residual $r$-dependent term which must be added to the 
non-leading terms computed in \cite{ps}. The sum of all these terms is
then minimized to find the favored position of the probe. The modification
of $detG_\parallel$ is the major thermal effect on the probe calculation.
Thermal modification of other quantities is considered in later sections of
this paper and produces corrections of higher order in $r_H^2/r^2$ which can
be ignored when the 5-brane is far from the horizon. The near-horizon
effect of these corrections will be included qualitatively here
and then supported by the work of later sections. We now proceed to implement
the probe calculation just outlined.

At finite temperature, the  contribution to the 
action from the Born-Infeld term in \eref{d5act} which is of leading order 
in the expansion parameter $n/\sqrt{gN}$ is
\begin{eqnarray}\label{leadBI}
S_0 \* -\frac{2\pi\a'\mu_5}{g}\int d^6\xi\left(-\det(\Gp)\right)^{1/2}
\left(\det(F_{\it2})\right)^{1/2}\nonumber\\
\* -\frac{\mu_5}{gn}4\pi^2\a'\frac{n^2r^4}{R^4}V\sqrt{1-r_H^4/r^4},
\end{eqnarray}
where $r_H$ is the position of the horizon and $V$ is the volume of
${\bf R}^4$.  
The leading contribution from the Chern-Simons
term is
\begin{equation}\label{CS0}
 2\pi\alpha'\mu_5\int {\cal F}_{\it 2} \wedge C_{\it 4}=
+\frac{\mu_5}{g}4\pi^2\a'\frac{nr^4}{R^4}V.
 \end{equation}
Dominant thermal corrections are absent since the metric does not appear
in this term.
Further only the parallel components of the Ramond-Ramond gauge potential
contribute to the integral, so we have set $C_{\it4}=\chi_{\it4}.$ 
The zero temperature first order correction to the Born-Infeld term plus
the contribution of the $C_{\it 6}$ plus the order $m^2$ term required by
supersymmetry is given in \cite{ps} as
\begin{equation}\label{BI1CS1}
\Delta S = -\frac{2\mu_5}{g\a'n}r^2(r-r_0)^2V,
\end{equation}
where
\begin{equation}
r_0 = \pi |m| n\a'.
\end{equation} 
Adding the terms in \eref{leadBI}, \eref{CS0} and \eref{BI1CS1} leads to the 
expression
\begin{equation}\label{d5pot}
-\frac{S}{V} = \frac{2\mu_5}{g\a'n}\left(\frac{\pi n^2}{2gN}r^4(
\sqrt{1-r_H^4/r^4}-1) + r^2(r-r_0)^2\right).
\end{equation}

This result for the free energy of the probe is valid if $r>>r_H$. Note 
that for $r_H=0$ \eref{d5pot} has a minimum at $r=r_0$ and a maximum at
$r=r_0/2.$ We now discuss thermal effects in this region of $r.$ 
For $r_H <<r$ the minimum and maximum shift by
\begin{equation}\label{minshift}
\Delta r_{min} = -\frac{\pi n^2 r_H^8}{2gNr_0^7},
\end{equation}
\begin{equation}\label{maxshift}
\Delta r_{max} = +\frac{32\pi n^2 r_H^8}{gNr_0^7},
\end{equation}
Therefore, as $r_H$ is increased, the maximum and minimum are pushed toward
each other, and the minimum disappears when $\Delta r_{max}\sim r_0$. This
occurs at the critical horizon size $r_H\sim (gN/n^2)^{1/8}r_0$ which is
well inside of $r_0$ and thus within the region of validity of the calculation.

To understand what happens closer to the horizon we must incorporate 
thermal effects on the second term in \eref{d5pot}. One effect is that the
$r^4$ term in \eref{BI1CS1} should be
multiplied by $\sqrt{1-r_H^4/r^4}$ since it comes from the Born-Infeld part of
\eref{d5act}. There are other effects from the 3-form perturbation and its 
back reaction on the metric and 5-form.  The linear term in $r_0$ comes
 from the Chern-Simons term 
\begin{equation}\label{horcomp}
\mu_5\int(C_{\it6}-B_{\it2}\wedge C_{\it4}),
\end{equation}
and goes to zero at the horizon due 
to a conspiracy between $C_{\it6}$ and
$B_{\it2}$.
The $r_0^2$ term has contributions from both the Born-Infeld 
and Chern-Simons parts of the probe action.  The Born-Infeld contribution is 
again  suppressed
by the factor $\sqrt{1-r_H^4/r^4}$ , but the Chern-Simons term also
vanishes since the order $m^2$ corrections to $C_{\it4}$ approach zero. 
These statements will be verified to order $m^2$ in sections 3 and 4.

The conclusion is therefore that the 5-brane has lower free energy at 
the horizon, so that the minimum near $r=r_0$ is a metastable state even
for small finite $r_H$ and is a degenerate vacuum state only
 if $r_H=0$.  
There is 
still another thermal effect which we discuss in section 6. 
Due to the back reaction of the 3-form
perturbation on the metric the horizon size itself shrinks as a function
of $m^2$ for fixed black hole temperature. Presumably $r_H$ vanishes at
some finite temperature $T_0$ of order $T_0\sim m$. This value would
then be a critical temperature for the probe calculation. In the low
temperature phase $T<T_0$ the 5-brane is stabilized near $r=r_0$; in
the high temperature phase $T>T_0$ it retreats to the horizon. Work is
continuing to verify this suggested picture.

Standard no-hair theorems should imply that the D5 brane at the horizon is
equivalent to $n$ D3 branes, since there is no net D5 brane charge.  At the
horizon, there is no contribution from the Born-Infeld action.  Hence, one
would need to verify that the actions coming from the Chern-Simons pieces
are equivalent.  The difference between the two actions is proportional to 
\eref{horcomp}.  As previously stated, the order $m^2$ contribution
will be shown to
be zero at the horizon.

The generalization of the previous analysis to NS5 probe branes with 
$p$ units of D3 brane
charge is straightforward. 
To proceed, one describes the D3 background using
the $S$-dual description
\begin{equation}
\tau' = \frac{a\t+b}{c\t+d}
\end{equation}
with field transformations
\begin{eqnarray}
g' \* g|M|^2\qquad\qquad G'_{MN} = G_{MN}|M|\qquad\qquad C_{\it4}=C_{\it4}
\nonumber\\
G_{\it3}'\* G_{\it3}\qquad\qquad B_{\it6}'-\t'C_{\it6}'=(B_{\it6}-\t 
C_{\it6})M^{-1}
\end{eqnarray}
where $M=c\t+d$.
Hence, the leading order term in \eref{d5pot}
is invariant under this transformation.
The first order terms have the same transformation as in \cite{ps},
\begin{eqnarray}\label{BI1CS1NS}
\Delta S \* \frac{2\mu_5}{g^3\a'p}r^2(r-\widetilde r_0)^2V.\nonumber\\
\widetilde r_0 \*  {\pi g|m| p\a'}
\end{eqnarray}
Hence, the total action that is minimized is 
\begin{equation}\label{ns5pot}
-\frac{S}{V} = \frac{2\mu_5}{g^3\a'p}\left(\frac{\pi gp^2}{2N}r^4(
\sqrt{1-r_H^4/r^4}-1) + r^2(r-\widetilde r_0)^2\right).
\end{equation}

Finally we point out that a more precise analysis in the Appendix shows that
the metastable D5 probe solution near $r=r_0$ disappears above the temperature
\begin{eqnarray}\label{Tbetterapp}
&&T \approx \kappa \left(\frac{n^2}{4\pi gN}\right)^{3/8}|m|\\
&& \kappa = \frac{1}{16}\left(\frac{54}{\pi^2}\right)^{1/8}\left(
3351797+171(2659)\sqrt{57}\right)^{1/8}
\approx  0.5522\nonumber
\end{eqnarray}
Likewise, for NS5 brane probes one finds that the solution disappears at
the temperature 
\begin{equation}\label{ns5better}
T\approx \kappa \left(\frac{gp^2}{4\pi N}\right)^{3/8}|m|,
\end{equation}

Hence these metastable  probe minima
 can survive well above the critical temperature,
at least for those probes that satisfy the probe condition $n^2>>gN$. 
This contrasts with the weak coupling analysis. 
At weak coupling, the presence of temperature introduces the
effective mass term $T^2\tr(\phi^2)$ to the effective potential.  Hence,
there are metastable Higgs vacua here as well.  However, in this case
these local minima are washed out when $T\sim m$.  

What is intriguing about these metastable vacua is that one could have
solutions where only some of the D3 brane charge is outside the horizon.
The corresponding vacuum is a metastable state where say part of the $SU(N)$
gauge group has been Higgsed, 
but the rest is unbroken and unconfined.  
Likewise, we can have metastable vacua where part of the gauge group
is confined, but the rest is unbroken and unconfined.

\section{The 3-Form Perturbation}

As in \cite{ps} the first step in the study of thermal effects is to obtain
linear perturbations of $H_{\it3}$  and $\tilde{F}_{\it3} = F_{\it3} -CH_{\it3}$  which are dual to
fermion mass terms in the boundary gauge theory. Specifically we must solve
the linearized equations of motion and Bianchi identities
in the background \eref{metric}, \eref{dil}--\eref{f5}:
\begin{eqnarray}\label{eom}
d*\tilde{F}_{\it3}\*F_{\it5}\wedge H_{\it3}\\
d*(\frac{1}{g}H_{\it3}-gC\tilde{F}_{\it3})\*-gF_{\it5}\wedge F_{\it3}\\
d\tilde{F}_{\it3}\*0=dH_{\it3}\
\end{eqnarray}
Expressing the fields as the complex combinations
\begin{eqnarray}\label{g3t}
G_{\it3}\*F_{\it3}-\tau H_{\it3}\\
\tau \* C +i/g,
\end{eqnarray}
the equations of motion in \eref{eom} can be recast into the compact form
\cite{ps}
\begin{eqnarray}\label{geom}
d*G_{\it3} +igG_{\it3}\wedge F_{\it5}\*0\\
dG_{\it3}\*0\
\end{eqnarray}

The black hole metric affects only the radial dependence of $G_{\it3}$, 
the dependence
on angles must be that of the lowest spherical harmonic on $S^5$ in 
order to be a
source for the fermion mass in the {\bf 10} or $\OL{bf{10}}$ 
representation of $SO(6)$.
Therefore we postulate that $G_{\it3}$ is the combination
\begin{equation}\label{ansatz}
G_{\it3}= \alpha(r)T_{\it3} + \beta(r) V_{\it3}\
\ee
of the same 3-forms used in\cite{ps}. The 3-forms $T_{\it3}$ and $V_{\it3}$ are 
constructed
from a constant antisymmetric tensor $T_{mnp}$ which is either self-dual or
anti-self-dual corresponding to the {\bf 10} or $\OL{\bf{10}}$ 
representations,
respectively. 
These forms are simply
written in terms of Cartesian coordinates $y^m, m=1,2,\cdots,6$ in
the space perpendicular to the 3-branes, with $r^2 = y^my^m$:
\begin{eqnarray}\label{tv}
T_{\it3}\*\frac{1}{3!} T_{mnp}dy^m\wedge dy^n\wedge dy^p\nonumber\\
S_{\it2} \*\frac{1}{2} T_{mnp}y^m dy^n\wedge dy^p\\
V_{\it3}\*d(\ln r)\wedge S_{\it2}\nonumber
\end{eqnarray}
One can easily derive the properties
\begin{eqnarray}\label{props}
*_6T_{\it3}\*=\pm i T_{\it3}\nonumber\\\  
*_6V_{\it3} \*\pm i (T_{\it3}-V_{\it3})\nonumber\\\
dT_{\it3}\*0\\  
dS_{\it2}\*3T_{\it3}\nonumber\\    
dV_{\it3}\*-3d(\ln r)\wedge T_{\it3}\nonumber
\end{eqnarray}
where $*_6$ is the Poincare duality operation in flat $ R^6$.

We now proceed to solve the equations \eref{geom}. Substitution of the ansatz 
\eref{ansatz} in
the Bianchi identity readily gives
\begin{equation}\label{bia}
dG_{\it3}= (\alpha'(r) -3\beta(r)/r) dr\wedge T_{\it3}=0\
\ee
which determines the relation $\beta(r) = r \alpha'(r)/3$. 
One can then directly show that 
\begin{equation}\label{Gclosed}
G_{\it3}~=~\frac{1}{3}d(\a(r)S_{\it2}).
\end{equation}
The equation of
motion in \eref{geom} contains the duality operation with respect to the 
full metric of \eref{metric},
\begin{equation}\label{fulldual}
*G_{\it3} = \frac{\sqrt{-g}}{3!} \epsilon_{mnpqst} g^{qq'}g^{ss'}g^{tt'}
G_{q's't'} dy^m\wedge dy^n\wedge dy^p\wedge gZ(r)\chi_{\it4} \
\end{equation}
where $g_{mn}$ is the transverse piece of the metric of \eref{metric} 
rewritten in terms of $y^m$, 
\begin{equation}\label{gcart}
g_{mn}~=~\left(\delta_{mn}-\frac{y^my^n}{r^2}\right)+f^{-1}(r)\frac{y^my^n}{r^2}.
\end{equation}
When compared to the case of zero temperature $(r_H=0)$,
one sees that the radial components
of $G_{\it3}$ dualize with an extra Schwarzschild factor $f(r)$ coming
from $g^{rr}$.

To handle this simply we split the forms $T_{\it3}$ and $V_{\it3}$ into radial and
angular parts, using
\begin{equation}
dy^m \equiv y^m d(\ln  r) + \omega^m\
\end{equation}
where $\omega^m$ is a set of angular 1-forms satisfying
\begin{equation}
d\omega^m = - dy^m \wedge d(\ln  r)\
\end{equation}
Then we have
\begin{equation}
T_{\it3} = \hat{T}_{\it3}  + \tilde{T}_{\it3}\
\end{equation}
with
\begin{eqnarray}
\hat{T}_{\it3} \* \frac{1}{2} T_{mnp} y^m d(\ln  r)\wedge \omega^n \wedge \omega^p
\nonumber\\
         \* \frac{1}{2} T_{mnp} y^m d(\ln  r)\wedge dy^n \wedge dy^p\nonumber\\
         \* V_{\it3}\\
\tilde{T}_{\it3}\* \frac{1}{3!} T_{mnp}\omega^m\wedge \omega^n\wedge \omega^p
\nonumber 
\end{eqnarray}
Note that $V_{\it3}= d(\ln r)\wedge S_{\it2}=\hat{V}_{\it3}$ has no angular part. 
The differentials are
\begin{equation}\label{diffs}
d\hat{T}_{\it3}~=~-3d(\ln  r)\wedge \tilde{T}_{\it3}\qquad\qquad d\tilde{T}_{\it3}~=~
 d\hat{V}_{\it3}\
\end{equation}
Under flat $*_6$ duality, we have 
\begin{equation}
*_6\left(\hat{T}_{\it3} + \tilde{T}_{\it3}\right) = \pm i(\hat{T}_{\it3}+\tilde{T}_{\it3})
\end{equation}
so that $*_6\hat{T}_{\it3} = \pm i \tilde{T}_{\it3}$ and $*_6\tilde{T}_{\it3} =\pm 
i\hat{T}_{\it3}$.
Under the duality \eref{fulldual}
of the 10-dimensional metric \eref{metric}, one thus has 
\begin{eqnarray}
*\hat{T}_{\it3}\*gf(r)*_6 \hat{T}_{\it3} \wedge \chi_{\it4}= \pm igf(r) \tilde{T}_{\it3}\wedge 
\chi_{\it4}\nonumber\\
 {*} \tilde{T}_{\it3}\*g*_6 \tilde{T}_{\it3} \wedge \chi_{\it4}=
\pm ig \hat{T}_{\it3}\wedge \chi_{\it4}\
 \end{eqnarray}
For $V_{\it3}$ \eref{props} implies 
\begin{equation}
*V_{\it3}~=~*\hat{T}_{\it3}=\pm igf(r)\tilde{T}_{\it3}\wedge\chi_{\it4}\
\end{equation}
Combining these duality relations with the ansatz \eref{ansatz}, we find 
\begin{eqnarray}
*G_{\it3}\*\alpha(r)*(\hat{T}_{\it3}+\tilde{T}_{\it3}) + \beta(r) *\hat{V}_{\it3}\nonumber\\
    \* \pm i[\alpha(r)(f(r)\tilde{T}_{\it3} +\hat{T}_{\it3}) 
+\beta(r) f(r) \tilde{T}_{\it3}]\wedge\chi_{\it4}\
\end{eqnarray}
Finally we note that 
\begin{equation}
G_{\it3}\wedge F_{\it5}~=~ G_{\it3}\wedge (d\chi_{\it4} + *d\chi_{\it4})~=~-d(G_{\it3}\wedge \chi_{\it4})
\end{equation}
where we have used the fact that $G_{\it3}\wedge*d\chi_{\it4} =0$ and $dG_{\it3}$ =0. 

Thus the 
equation of motion can be simply rewritten as 
\begin{eqnarray}\label{eom2}
d[*G_{\it3} -iG_{\it3}\wedge \chi_{\it4}] \* i\ d\left(\pm \left[\alpha(r)(f(r)\tilde{T}_{\it3}
+\hat{T}_{\it3})
 +
\beta(r) f(r) \tilde{T}_{\it3} -(\alpha(r) T_{\it3}+\beta(r) \hat{T}_{\it3})\right]\wedge 
\chi_{\it4}\right)\nonumber\\
\*0\
\end{eqnarray}
Using 
\begin{equation}
d\chi_{\it4}= (Z'/Z)dr\wedge \chi_{\it4}=-4d(\ln r)\wedge \chi_{\it4}
\end{equation}
 and \eref{diffs}
it is straightforward 
to work out the differential, substitute $\beta = r \alpha'/3$, and obtain 
the differential equation for $\alpha(r)$:
\begin{equation}
rf(r) \alpha''(r) + (11f(r)+rf'(r)) \alpha'(r) +3(f'(r)+7f(r)/r+\gamma r)\alpha(r)=0\
\end{equation}
where $\gamma=-7$ in the self dual case $*_6T_{\it3}= iT_{\it3}$ and $\gamma=1$ in the
anti-self-dual case $*_6T_{\it3}= -iT_{\it3}$. We now restrict to the latter case for
which the fluctuation $G_{\it3}$ is in the $\OL{\bf{10}}$ representation 
dual to the
fermion mass term $tr\lambda\lambda$ in the \noneplus\ field theory. 
Inserting the specific form of $f(r)$ and
making the substitutions $\alpha(r)= F(u)/r^4$ with $u=r_H^2/r^2$, one finds
the Legendre equation
\begin{equation}
F''(u) -\frac{2u}{1-u^2} F'(u)- \frac{1}{4(1-u^2)}F(u)=0.
\end{equation}
The general solution is 
\begin{equation}\label{legendresol}
\alpha(r) = \frac{1}{r^4} [a P_{-\frac{1}{2}}(r_H^2/r^2) +
   bQ_{-\frac{1}{2}}(r_H^2/r^2)]
\end{equation}
Only the first term is regular at the horizon, so we set $b=0$ and observe
that this solution has the expected \cite{ps} behavior $1/r^4$ as 
$r \to\infty$ associated with the fermion bilinear $tr{\lambda\lambda}$ which
is an operator of scale dimension 3.  The coefficient $a$ is set by comparing
\eref{legendresol} to the zero temperature result \cite{ps}, 
giving the solution
\begin{equation}\label{normalized}
\a(r)= -\frac{3\sqrt{2}\Gamma(3/4)^2}{g\sqrt{\pi}}\frac{R^4}{r^4}
P_{-\frac{1}{2}}(r_H^2/r^2).
\end{equation}
We also observe that there is no regular solution which approaches the boundary
at the rate $1/r^6$ associated with a vacuum expectation value
$\langle\tr\lambda\lambda\rangle$.

In the zero temperature case, $G_{\it3}$ is singular at the origin.  It was thus
necessary to insert D5 or NS5 branes wrapping two-spheres in order to
interpret the  singularity.  In this case, where there is a black hole with a 
finite Schwarzschild radius, all singularities are hidden behind the horizon.
Hence, one is no longer compelled to insert 5 branes.  From the gauge theory
point of view, one should interpret this as being in the high temperature phase
where the system acts like a finite temperature \nfour\ theory.  Only
when $r_H$ shrinks to zero will the solution for $\a(r)$ be singular, and
thus necessitate the insertion of 5 branes.

Let us now turn to the 6-form dual potentials.  
Making use of the Bianchi identity
and the relation of $V_{\it3}$ to $T_{\it3}$, we can rewrite the expression for $G_{\it3}$
in \eref{ansatz} as 
\begin{equation}\label{newG_3}
G_{\it3}~=~\a(r)\widetilde T_{\it3} + (\a(r)+(1/3)r\a'(r))\hat T_{\it3}.
\end{equation}
The equation \eref{eom2} implies that $*G_{\it3}-i G_{\it3}\wedge \chi_{\it4}$ is a closed
7-form.  Hence, it can be written as \cite{ps}
\begin{eqnarray}\label{cl7form}
*G_{\it3}-i G_{\it3}\wedge \chi_{\it4} \*\frac{g}{i} d(B_{\it6}-\tau C_{\it6})\nonumber\\
                        \*\frac{g}{i} d\left(\frac{-i}{3}
\frac{\a(r)+f(r)(\a(r)+r\a'(r)/3)}{Z}S_{\it2}\right)dx^0\wedge dx^1\wedge dx^2\wedge
dx^3,
\end{eqnarray}
where the last line is derived using the relations in \eref{props}.  If we assume that $C=0$, and hence $\tau$ is imaginary, then $C_{\it6}$ is given by
\begin{equation}\label{C6eq}
C_{\it6}=-\frac{g}{3}\left(\a(r)+f(r)(\a(r)+r\a'(r)/3)\right)\Im S_{\it2}\wedge\chi_{\it4}.
\end{equation}

It is instructive to compare this result for $C_{\it6}$ to
the NS-NS 2-form.  Using  \eref{g3t} and \eref{Gclosed}, one finds that
\begin{equation}\label{B2eq}
B_{\it2}=-\frac{g}{3}\a(r)\Im S_{\it2}.
\end{equation}
Since $f(r)=0$ at the horizon, we learn that $C_{\it6}=B_{\it2}\wedge\chi_{\it4}$ at $r=r_H$.
This confirms the claim in the previous section that a D5 brane with D3 brane
charge $n$ has the same action at the horizon as $n$ D3 branes.

\section{Back reaction of the perturbation on the dilaton}

In this section we compute the order $m^2$ corrections to the dilaton due
to the temperature dependent $G_{\it3}$ perturbation of the previous section.
The modified dilaton actually contributes only higher order corrections
to many of the physical effects we are interested in, but the calculation is
simpler than for the 5-form and metric and serves as a prototype. The
result is of interest for $T=0$, since the order $m^2$ correction to the
dilaton, and hence the gauge coupling, is independent of the vacuum state.

We must solve the equation
\begin{equation}\label{dilaton}
D^MD_M \Phi =-\frac{1}{12} H_{MNP}H^{MNP} +\frac{g^2}{12} \tilde{F}_{MNP}
   \tilde{F}^{MNP},
\end{equation}
which was obtained from the more general dilaton equation of motion in
\cite{ps} by specializing dilaton and axion sources to their constant values in
\eref{dil} and \eref{axion}. 
The metric is that of \eref{metric} with perpendicular part rewritten
in terms of ``Cartesian'' coordinates $y^m$ as in \eref{gcart}. The source can
be expressed in terms of $G_{\it3}$ as 
\begin{eqnarray}\label{dil2}
D^MD_M \Phi \* \frac{g^2}{24} g^{mm'}g^{nn'}g^{pp'}
                             [G_{mnp} G_{m'n'p'} + c.c.]\nonumber\\
            \* \frac{g^2}{24 Z^{\frac{3}{2}}} 
[G_{mnp} G_{mnp} +\frac{3(f(r)-1)}{r^2}y^mG_{mnp} y^{m'}G_{m'np} + c.c.]
\end{eqnarray}
in which lower pairs of indices are summed over their 6 values. Using the
representation \eref{ansatz} and the fact that $y^mV_{mnp}=y^mT_{mnp}$, we 
obtain
\begin{eqnarray}\label{source1}
&&G_{mnp} G_{mnp} +\frac{3(f(r)-1)}{r^2}y^mG_{mnp} y^{m'}G_{m'np}= \nonumber\\
&&\qquad\qquad\qquad\alpha^2(r) T_{pqr}T_{pqr} -\frac{3\a^2(r)
-3f(r)(\alpha(r)+r\a'(r)/3)^2}{ r^2}y^my^nT_{mpq}T_{npq}, 
\end{eqnarray}
where $\a(r)$  is the same functions derived in the previous
section.

We now need the specific form of the anti-self-dual tensor $T_{mnp}$ whose
non-vanishing complex components are \cite{ps} 
\begin{equation}\label{ttensor}
T_{\bar{p}\bar{q}r}=T_{p\bar{q}\bar{r}}=T_{\bar{p}q\bar{r}}=m \epsilon_{pqr}
\end{equation}
where $\epsilon_{pqr}$ is the Levi-Civita symbol in 3 (complex) dimensions.
By conversion to complex coordinates and reconversion, we obtain
\begin{eqnarray}
T_{pqr}T_{pqr}\* 0\nonumber\\
y^my^nT_{mpq}T_{npq} \* 2 m^2 r^2 Y(y^i/r)\\
Y(y^i/r)\*\frac{ (y^{1})^2+(y^{2})^2+(y^{3})^2-(y^4)^2-(y^5)^2-(y^6)^2}{r^2}\nonumber
\end{eqnarray}
The function $Y(y^i/r)$ is a harmonic polynomial for the {\bf 20'} 
representation
of $SO(6)$ and has ``eigenvalue'' $-12/r^2$ of the flat 6-dimensional 
Laplacian 
in the $y^i$. Thus the final equation for the lowest order dilaton perturbation
is
\begin{equation}\label{dil3}
D^MD_M \Phi =-\frac{m^2g^2}{2 Z^{\frac{3}{2}}}[\a^2(r)
-f(r)(\alpha(r)+r\a'(r)/3)^2] Y(y^i/r),
\end{equation}
where we have assumed that $m^2$ is real to prevent a  source term
for the Ramond-Ramond scalar field.

The next step is to find the specific form of the Laplacian. Since the
source depends only on $r$ and angles, we can drop $t$ and $\bf{x}$
derivatives. The solution must also be proportional to the harmonic $Y(y^i/r)$
so that angular derivatives just give the eigenvalue $-12$. Thus
\begin{equation}\label{lapl}
D^MD_M \Phi = D^MD_M \phi(r) Y(y^i/r)
            = \frac{r^2f(r)}{R^2}[\partial^2/\partial r^2 +(\frac{5}{r}+
\frac{f'}{f})\partial/\partial r) -\frac{12}{r^2f(r)}]\phi(r) Y(y^i/r)
\end{equation}

In the limit of zero temperature $(f(r)\rightarrow1)$ \eref{lapl} and 
\eref{dil3} combine to give the radial differential equation
\begin{equation}\label{zerodil}
(\frac{d^2}{dr^2} + \frac{5}{r}\frac{d}{dr}-\frac{12}{r^2})\phi(r) =
-\frac{8m^2R^4}{r^4}
\end{equation}
This equation has homogeneous
solutions $\phi(r) \propto r^2,r^{-6}$ which are the power laws of the
irregular and regular solutions for a field of scale dimension $\Delta=6$
in the {\bf 20'} representation of $SO(6)$. This is the second Kaluza-Klein
excitation of the dilaton/axion \cite{krvn}. The inhomogeneous solution
is
\begin{equation}\label{dil4}
\phi(r) = \frac{m^2R^4}{2r^2}.
\end{equation}

Presumably this nontrivial behavior for the dilaton
 describes the scale dependence of the field theory gauge
coupling due to the introduction of mass for the chiral multiplets.
However, we do not clearly understand the angular dependence of
$\Phi= \phi(r) Y(y^i/r)$ and thus cannot give a more precise
interpretation.  For a softly broken \nfour\ to \ntwo\ theory one finds
angular dependence for the effective couplings on the coulomb branch
\cite{Pilch:2000ue}.   However, for \nfour\ broken to \none\ there is
no coulomb branch, so the Wilsonian interpretation seems less clear.

In this connection it is interesting to note that for the ($T=0$)
solution $\alpha(r) \sim 1/r^6$, which is dual to the field theory vev
$\langle\tr(\lambda\lambda)\rangle$, 
the source term in \eref{dil3} vanishes. Thus the
dilaton does not run in a situation where no masses are turned on
\cite{Petrini:1999qa,Pilch:2000ue}.

In \cite{ps} an interpolating function for the dilaton was given that
followed a 5-brane geometry in the IR to a 3-brane geometry in the UV.
For example, for a vacuum that corresponds to one D5 brane with D3 brane
charge $N$, the interpolating dilaton was given by
\begin{equation}\label{intdil}
e^{2\Phi}=g^2\frac{\rho_-^2}{\rho_-^2+\rho_c^2},
\end{equation}
where
\begin{eqnarray}\label{rhoeqs}
\rho_-^2\* {y_1}^2+{y_2}^2+{y_3}^2
+\left(\sqrt{{y_4}^2+{y_5}^2+{y_6}^2}-r_0\right)^2\nonumber\\
\rho_c\* \frac{2gr_0\a'}{R^2}
\qquad\qquad
r_0=\pi\a' m N.
\end{eqnarray}
If we expand this for large $r$ we find that the order $m^2$ dilaton correction
is
\begin{equation}\label{asyintdil}
\delta \Phi(y_i) = -\frac{m^2R^4}{8r^2}.
\end{equation}
This is an $SO(6)$ singlet, 
hence the first order correction in the interpolating
dilaton is not quite
 consistent with the supergravity equations of motion.  If one
were to have considered an NS5 brane, then one would have found 
\eref{asyintdil} but with the opposite sign.  
In fact, if one were to solve the supergravity equations but with
5-brane sources inserted, one would find that their effects are of higher
order in $m$.  The order $m^2$ term should be independent of the 
particular vacuum the \noneplus\ theory is in.

At finite temperature the equation in \eref{dil3}
is more complicated but tractable. Using
the variable $z=Z(r)=R^4/r^4$ and setting $k=r_H^4/R^4$,
we find the radial differential operator
\begin{equation}
[\partial^2/\partial r^2 +(\frac{5}{r}+
\frac{f'}{f})\partial/\partial r -\frac{12}{r^2f(r)}] \phi(r)=
16\frac{z^{\frac{5}{2}}}{R^2}[\frac{d^2}{dz^2}-\frac{k}{1-kz}\frac{d}{dz}-
\frac{3}{4z^2(1-kz)}]\phi(z)
\end{equation}
which is essentially hypergeometric. Its zero modes are:
\begin{eqnarray}\label{dilhom}
\phi_{1}(z) \* z^{\frac{3}{2}} F(\frac{3}{2},\frac{3}{2};3;kz)
           \rightarrow z^{\frac{3}{2}}\\ 
\phi_{2}(z) \* z^{\frac{3}{2}} F(\frac{3}{2},\frac{3}{2};1;1-kz)
           \rightarrow \frac{4}{\pi k^2 z^{\frac{1}{2}}}\
\end{eqnarray}
whose Wronskian is
\begin{equation}\label{wron}
W(\phi_1,\phi_2) = -\frac{8}{\pi k^2(1-z)}.\
\end{equation}
The limiting form of the solutions at the boundary $(z\rightarrow 0)$ is also
given in \eref{dilhom}. At the horizon $(z\rightarrow 1/k)$, the function
$\phi_1(z)$ has a logarithmic singularity while $\phi_2(z)$ is regular.

The finite temperature equation with source \eref{dil3} can be rewritten as
\begin{eqnarray}\label{inhom}
&&[\frac{d^2}{dz^2} -\frac{k}{1-kz} \frac{d}{dz} -\frac{3}{4z^2(1-kz)}]\phi(r)
= \frac{S(z)}{1-kz}\\
&&S(z)\equiv-\frac{m^2g^2R^2}{32z^{\frac{7}{2}}}\left[\a^2(z)
-f(z)(\alpha(z)-4\a'(z)/3)^2\right].
\end{eqnarray}
The method of variation of parameters gives the solution
\begin{equation}\label{phisol}
\phi(r) = -\frac{\pi k^2}{8}\left[\phi_1(z) \int_{z}^{1/k}dz \phi_2(z)S(z)
     +\phi_2(z) \int_{0}^{z}dz \phi_1(z)S(z)\right]
\end{equation}
which is regular on the horizon and vanishes on the boundary at the same 
rate as \eref{dil4}. This solution
does not appear to be integrable in closed form.\\

\section{Back reaction to the metric and self dual 5-form}

In this section we compute leading order corrections to the metric and 
self dual 5-form when the field $G_{\it3}$ is turned on.
The Einstein equations are
\begin{equation}\label{einstein}
R_{MN}-\half g_{MN}R = \frac{g^2}{48}F_{MPQRS}{F_{N}}^{PQRS}+\frac{g^2}{8}
(G_{MPQ}{G_{N}^*}^{PQ}+c.c)-\frac{g^2}{24}g_{MN}G_{PQR}{G^*}^{PQR},
\end{equation} 
where the contribution of the dilaton is ignored since it affects only higher
order terms.

Let us consider the $G_{\it3}$ terms first.  Using the results in \eref{g3t}, 
\eref{Gclosed} and \eref{gcart}, it is straightforward to show that
\begin{eqnarray}\label{GmGn}
&&\frac{1}{8}G_{mpq}{G_n^*}^{pq}+c.c.=\nonumber\\
&&\qquad\frac{|T_{\it3}|^2}{24Z}\left(
\a(r)^2\left(3I_{mn}- 2W_{mn}\right) +
(\a(r)+r\a'(r)/3)^2\left(\frac{y^my^n}{r^2}+2f(r)(W_{mn}+I_{mn})\right)\right),
\end{eqnarray}
where $\a(r)$ is the same as in \eref{normalized},
$|T_{\it3}|^2=T_{pqr}{T^*}^{pqr}$ and 
\begin{eqnarray}\label{Wdef}
I_{mn} \* \frac{1}{5}\left(\de_{mn}-\frac{y^my^n}{r^2}\right)\nonumber\\
W_{mn}\* \frac{3}{|T_{\it3}|^2}\left(T_{mpk}T^*_{npl}\frac{y^ky^l}{r^2}+c.c\right)
-I_{mn}.
\end{eqnarray}
Using the fact that 
\begin{equation}
\half(T_{mpq}T^*_{npq}+c.c)=\frac{1}{6}\de_{mn}|T_{\it3}|^2,
\end{equation}
one finds that $W_{mn}$ is traceless.  We also see that $W_{mn}$ is
strictly an angular tensor since $y^mW_{mn}=0$ because of the antisymmetry
of $T_{mnp}$.  

Taking the trace of the expression in \eref{GmGn} gives
\begin{equation}\label{G2trace}
G_{mpq}{G^*}^{mpq}
=\frac{|T_{\it3}|^2}{2Z^{3/2}}\left(\a(r)^2+(\a(r)+r\a'(r)/3)^2\right).
\end{equation}
Hence the contribution to the energy-momentum tensor from the $G_{\it3}$ kinetic 
terms is 
\begin{eqnarray}\label{Gemt}
&&\frac{g^2}{8}G_{mpq}{G_n^*}^{pq}+c.c.-\frac{g^2}{24}g_{mn}G_{pqr}{G^*}^{pqr}=
\nonumber\\
&&\qquad\frac{g^2|T_{\it3}|^2}{48Z}(\a(r)^2-f(r)(\a(r)+r\a'(r)/3)^2)\left(I_{mn}
-4W_{mn}-f^{-1}(r)\frac{y^my^n}{r^2}\right).
\end{eqnarray}
for the orthogonal components and
\begin{eqnarray}
-\frac{1}{24}g_{ij}G_{pqr}{G^*}^{pqr}\*
-\frac{|T_{\it3}|^2}{48Z^2}\left(\a(r)^2+(\a(r)+r\a'(r)/3)^2\right)\eta_{ij}
\label{Gpemt}\\
-\frac{1}{24}g_{00}G_{pqr}{G^*}^{pqr}\*
\frac{|T_{\it3}|^2}{48Z^2}\left(\a(r)^2+(\a(r)+r\a'(r)/3)^2\right)f(r)\label{Gpemt2},
\end{eqnarray}
for the space-time components.

We next compute the corrections to 
$\widetilde F_{\it5}=F_{\it5}-\half (C_{\it2}\wedge H_{\it3}-B_{\it2}\wedge F_{\it3})$, 
which in terms of $G_{\it3}$ is 
\begin{equation}\label{tF5F5}
\widetilde F_{\it5}= F_{\it5}+\frac{ig}{4}(\psi\wedge G_{\it3}^*-\psi^*\wedge G_{\it3}),
\end{equation}
where $\psi=\frac{1}{3}\a(r)S_{\it2}$.  The equation for motion for 
$\widetilde F_{\it5}$
is 
\begin{equation}\label{F5eom}
\tF_{\it5}=*\tF_{\it5}.
\end{equation}
There are two correction terms that need to be added to the lowest order term
\begin{equation}\label{F5zero}
F_{\it5}^0=d\chi_{{\it4},0}+*d\chi_{{\it4},0}
\end{equation}
in order to satisfy \eref{F5eom}, where $\chi_{{\it4},0}$ is the 4-form field
in \eref{chieq}.
First, since there are corrections to the metric, $\chi_{{\it4},0}$ 
is no longer
harmonic and so the expression in 
\eref{F5zero} is not closed.  We will return to this point
once we discuss corrections to the metric. Second, the
improvement term in \eref{tF5F5} is not self-dual,  so it is 
necessary to add another correction term to $F_{\it5}$ to compensate for
this term.  A good guess is to add the dual of 
the improvement term to $F_{\it5}$, which would make the sum of this term and
the improvement term manifestly self-dual.
However, one first needs to verify that
the dual is closed.  To see this,
we note that 
\begin{eqnarray}\label{G3eq}
G_{\it3}\*\frac{1}{3}\a'(r)dr\wedge S_{\it2} + \a(r) T_{\it3}\nonumber\\
        \*\left(\frac{1}{3}\a'(r)dr+r^{-1}\a(r)\right)\wedge S_{\it2} +
\frac{1}{3} r^3\a(r)d(r^{-3}S_{\it2}),
\end{eqnarray}
where the second term on the rhs of the second line in \eref{G3eq} has no
radial components.  Therefore, we find that
\begin{equation}\label{G3eq2}
\frac{ig}{4}(\psi\wedge G_{\it3}^*-\psi^*\wedge G_{\it3})=
\frac{ig}{12}\a^2(r)\left(S_{\it2}\wedge T^*_{\it3}-S^*_{\it2}\wedge T_{\it3}
\right),
\end{equation}
with no radial components.  Therefore, the dual of \eref{G3eq2} is
\begin{eqnarray}\label{G3eq3}
&&\frac{ig\a^2(r)}{12Z^2}*_6\left(S_{\it2}\wedge T^*_{\it3}-
S^*_{\it2}\wedge T_{\it3}\right)\wedge dx^0\wedge dx^1\wedge dx^2\wedge dx^3=
\nonumber\\
&&\qquad\qquad -\frac{g\a(r)^2|T_{\it3}|^2}{72Z^2}rdr\wedge 
dx^0\wedge dx^1\wedge dx^2\wedge dx^3,
\end{eqnarray}
which only has a radial component in the orthogonal directions.  Since
$\a(r)$ and $Z$ only have $r$ dependence, the expression in \eref{G3eq3}
is clearly closed.

There is an $m^2$ correction coming from the cross term of
the improvement term and its dual with $F_{\it5}^0$.  For the angular, {\it 
i.~e.} 
magnetic, components,
the contribution of this cross term to \eref{einstein} is
\begin{eqnarray}\label{crossang}
&&-\frac{ig^2\a^2(r)rZ'}{288Z^2}\epsilon^6_{ijkmlq}\frac{y^iy^p}{r^2}
\left(3\left(
T_{pjk}T^*_{nlq}-c.c.\right)-2\left(T_{pjn}T^*_{klq}-c.c.\right)\right)=
\nonumber\\
&&\qquad\qquad -\frac{g^2\a^2(r)|T_{\it3}|^2}{36Z}\left(\de_{mn}-\frac{y^my^n}{r^2},
\right)
\end{eqnarray}
where we have used the fact that $*_6 T_{\it3}=-i T_{\it3}$. 
It then follows that the electric cross term piece is
\begin{equation}\label{crossel}
\frac{g^2\a^2(r)|T_{\it3}|^2}{36Z^2} g_{MN},
\end{equation}
where $g_{MN}$ is the $AdS_5$ Schwarzschild metric given in \eref{metric}

We can now combine the results of \eref{Gemt}, \eref{Gpemt}, \eref{Gpemt2},
\eref{crossang}
and \eref{crossel} into one source term, $J_{MN}$, for the right-hand side of 
\eref{einstein}.
For the orthogonal components we have
\begin{eqnarray}\label{Jorth}
&&J_{mn}=-\frac{g^2|T_{\it3}|^2}{144Z}\left(\a^2(r)+3f(r)[\a(r)+r\a'(r)/3]^2\right)
\left(5I_{mn}-f^{-1}(r)\frac{y^my^n}{r^2}
\right)\nonumber\\
&&\qquad\qquad\qquad\qquad
-\frac{g^2|T_{\it3}|^2}{12Z}\left(\a^2(r)-f(r)[\a(r)+r\a'(r)/3]^2\right)
(W_{mn}+I_{mn}),
\end{eqnarray}
while the space-time components are given by
\begin{eqnarray}
J_{ij}\*+\frac{g^2|T_{\it3}|^2}{144Z}\left(\a^2(r)-3f(r)[\a(r)+r\a'(r)/3]^2\right)
\eta_{ij}\label{Jst1}\\
J_{00}\*\frac{g^2|T_{\it3}|^2}{144Z}\left(\a^2(r)-3f(r)[\a(r)+r\a'(r)/3]^2\right)
f(r)\label{Jst2}.
\end{eqnarray}

Upon inspection of \eref{Jorth} one sees that the right hand side consists of
three types of terms.  The radial term $y^my^n/r^2$ and the two angular terms
$I_{mn}$ and $W_{mn}$. Hence, the corrections to the transverse
metric components should be of this same type.  Likewise, \eref{Jst1}
indicates that the  spatial piece of the metric can have a correction
proportional to $\eta_{ij}$ while  
\eref{Jst2} indicates that  the temporal component 
has a correction proportional to $f(r)$.   Hence,
we can write down the following ansatz for the metric
\begin{eqnarray}\label{metcorr}
&&ds^2=-\left(Z^{-1/2}+h_0(r)\right)f(r)dt^2+\left(Z^{-1/2}+h_1(r)\right)\eta_{ij}dx^idx^j+\nonumber\\
&&\qquad\left[(5Z^{1/2}+p(r))I_{mn}+
f^{-1}(r)(Z^{1/2}+q(r))\frac{y^my^n}{r^2}+w(r)W_{mn}\right]dy^mdy^n.
\end{eqnarray}

Let us now return to the corrections to $F_{\it5}$ induced by metric 
corrections.  Let us write the 4-form 
$\chi_{\it 4}(r)=\chi_{{\it4},0}(r)+\chi_{{\it4},1}(r)$, where
 $\chi_{{\it4},1}(r)$ is assumed to have components in the 
four space-time directions,
\begin{equation}
\chi_{{\it4},1}(r)=\chi_1(r)dx^0\wedge dx^1\wedge dx^2\wedge dx^3 
\end{equation}
  Therefore, the dual of $d\chi_{\it4}(r)$ is
\begin{eqnarray}\label{chidual}
&&*(d\chi_{{\it4},0}(r)+d\chi_{{\it4},1}(r))=\frac{1}{5\!}\epsilon^6_{ijklmn}
(\det g_{||})^{-1}\sqrt{-det g}
g^{np}\frac{y^p}{r}\left(-\frac{-Z'}{gZ^2}+\chi_1'(r)\right)\nonumber\\
&&\qquad\qquad\qquad\qquad\qquad\qquad\qquad\times
 dy^i\wedge dy^j\wedge dy^k\wedge dy^l\wedge dy^m.
\end{eqnarray}
Up to  first order corrections this is
\begin{eqnarray}\label{chi1dual}
&&*(d\chi_0(r)+d\chi_1(r))=*_0d\chi_0(r)+\nonumber\\
&&\qquad\qquad\left[-\frac{Z'}{2gZ^{3/2}}\left(
\frac{p(r)-q(r)}{Z}-h_0(r)-3h_1(r)\right)+\chi_1'(r)\right]*_6d(\ln r),
\end{eqnarray}
where $*_0$ refers to the dual in the background metric.  Hence, 
$\chi_{\it4}(r)$
is harmonic with respect to the new metric if the term in square brackets in
\eref{chi1dual} is zero.  Thus, we set
\begin{equation}\label{chi1eq}
\chi_1'(r)=\frac{Z'}{2gZ^{3/2}}\left(
\frac{p(r)-q(r)}{Z}-h_0(r)-3h_1(r)\right)
\end{equation}
Plugging in this value for $\chi_1(r)$, one finds that the leading order
corrections to
the stress tensor coming from the cross term of $d\chi_{{\it4},1}(r)$ and 
$d\chi_{{\it4},0}(r)$, denoted by $K_{MN}$,
are
\begin{equation}\label{F5metorth}
K_{mn}=\frac{(Z')^2}{4Z^{5/2}}\left\{-4p(r)I_{mn}
+w(r)W_{mn}
+[p(r)-q(r)]f^{-1}(r)\frac{y^my^n}{r^2}\right\}
\end{equation}
for the orthogonal components and  
\begin{eqnarray}
K_{ij}\*\frac{(Z')^2}{4Z^{5/2}}\left[p(r)-Zh_1(r)\right]\eta_{ij}
\label{F5metst1}\\
K_{00}\*-\frac{(Z')^2}{4Z^{5/2}}\left[p(r)-Zh_0(r)\right]f(r)
\label{F5metst2}
\end{eqnarray}
for the space-time components.

The next step is to compute the linear corrections to the Einstein tensor by
inserting these expressions for $J_{MN}$ and $K_{MN}$ into \eref{einstein}.
The space-time components  will have the form
\begin{eqnarray}
R_{00}-\half g_{00}R \*- N_0(r)f(r)\label{einst1}\\
R_{ij}-\half g_{ij}R \* N_1(r)\eta_{ij}\label{einst2}
\end{eqnarray}
while the orthogonal components
will again be of the form
\begin{equation}\label{ein1}
R_{mn}-\half g_{mn}R =
N_2(r)f^{-1}(r)\frac{y^my^n}{r^2}+ 
N_3(r)I_{mn}+N_4(r)W_{mn}.
\end{equation}

Computing the functions in \eref{einst1}, \eref{einst2} and \eref{ein1}
is an extremely tedious exercise.  We just state the results here.
We find that
\begin{eqnarray}\label{N0eq}
&&N_0(r)=\frac{1}{Z^{3/2}}\biggl[\frac{1}{2}f(r)p''(r)+
(1+3f(r))\frac{1}{r}p'(r)+(4+3f(r))\frac{1}{r^2}p(r)
-\frac{3}{2}f(r)\frac{1}{r}q'(r)\nonumber\\
&&\qquad\qquad\qquad
-(6+3f(r))q(r)
+\frac{3Z}{2}f(r)h_1''(r)+\frac{3Z}{2}(2-f(r))\frac{1}{r}h_1'(r)-
6Z\frac{1}{r^2}h_1(r)\\
&&\qquad\qquad\qquad
-4Z\frac{1}{r^2}h_0(r)\biggr]\nonumber\\
\label{N1eq}
&&N_1(r)=\frac{1}{Z^{3/2}}\biggl[\frac{1}{2}f(r)p''(r)+
2(1+f(r))\frac{1}{r}p'(r)+(6+f(r))\frac{1}{r^2}p(r)
\nonumber\\
&&\qquad\qquad\qquad
-\frac{1}{2}(2+f(r))\frac{1}{r}q'(r)-(8+f(r))q(r)
\nonumber\\
&&\qquad\qquad\qquad
+Zf(r)h_1''(r)+Z(4-3f(r))\frac{1}{r}h_1'(r)-
Z(12-4f(r))\frac{1}{r^2}h_1(r)\\
&&\qquad\qquad\qquad
+\frac{Z}{2}f(r)h_0''(r)+\frac{Z}{2}(6-5f(r))\frac{1}{r}h_0'(r)
-Z(6-4f(r))\frac{1}{r^2}h_0(r)\biggr]\nonumber
\\
\label{N2eq}
&&N_2(r)=\frac{1}{Z^{1/2}}\biggl[
(1+f(r))\frac{1}{r}p'(r)+2(2+f(r))\frac{1}{r^2}p(r)-10q(r)
\nonumber\\
&&\qquad\qquad\qquad
+\frac{3Z}{2}(2+f(r))\frac{1}{r}h_1'(r)-
Z(6+3f(r))\frac{1}{r^2}h_1(r)\\
&&\qquad\qquad\qquad
+\frac{3Z}{2}f(r)\frac{1}{r}h_0'(r)
-3Zf(r)\frac{1}{r^2}h_0(r)\biggr]\nonumber\\
\label{N3eq}
&&N_3(r)=\frac{1}{Z^{1/2}}\biggl[2f(r)f_1''(r)+
(8+5f(r))\frac{1}{r}f_1'(r)+(26+8f(r))\frac{1}{r^2}f_1(r)
\nonumber\\
&&\qquad\qquad\qquad
-5(1+f(r))\frac{1}{r}q'(r)-10(6+f(r))q(r)
\nonumber\\
&&\qquad\qquad\qquad
+\frac{15Z}{2}f(r)h_1''(r)+10Z(2-f(r))\frac{1}{r}h_1'(r)-
15Z(4-f(r))\frac{1}{r^2}h_1(r)\\
&&\qquad\qquad\qquad
+\frac{5Z}{2}f(r)h_0''(r)+5Z(3-2f(r))\frac{1}{r}h_0'(r)
-15Z(2-f(r))\frac{1}{r^2}h_0(r)\biggr]\nonumber
\\
\label{N4eq}
&&N_4(r)=\frac{1}{Z^{1/2}}\biggl[-\frac{1}{2}f(r)w''(r)
-\frac{1}{2}(4+5f(r))\frac{1}{r}w'(r)+(6-2f(r))\frac{1}{r^2}w(r)\biggr].
\end{eqnarray}

 Using equations
\eref{Jorth}--\eref{Jst2}, \eref{F5metorth}--\eref{F5metst2} and
\eref{N0eq}--\eref{N4eq} we can reduce \eref{einstein} to five coupled
inhomogenous equations.  We write these as
\begin{eqnarray}\label{Leqs}
L_0(z)\*\frac{1}{12}A(z)-\frac{1}{4}B(z\nonumber)\\
L_1(z)\*\frac{1}{12}A(z)-\frac{1}{4}B(z)\nonumber\\
L_2(z)\*\frac{1}{12}A(z)+\frac{1}{4}B(z)\\
L_3(z)\*-\frac{17}{12}A(z)-\frac{1}{4}B(z)\nonumber\\
L_4(z)\*-A(z)+B(z)\nonumber
\end{eqnarray}
where again $z=R^4/r^4$ and 
\begin{eqnarray}\label{ABeq}
A(z)\*\frac{g^2|T_{\it3}|^2R^2}{12z}\a^2(z)\nonumber\\
B(z)\*\frac{g^2|T_{\it3}|^2R^2}{12z}
\left(\a(z)-\frac{4}{3}z\a'(z)\right)^2(1-kz)=\s(z)(1-kz),
\end{eqnarray}
with $k=r_H^4/R^4$.  The $L_i(z)$ functions are linear combinations of the
metric components and are given by
\begin{eqnarray}
\label{L0eq}
&&L_0(z)=8(1-kz)z^2p''(z)-2(3-kz)zp'(z)+3(1-kz)p(z)\nonumber\\
&&\qquad\qquad
+6(1-kz)zq'(z)-3(3-kz)q(z)\\
&&\qquad\qquad
+24(1-kz)z^3h_1''(z)+12(2-3kz)z^2h_1'(z)-6zh_1(z)\nonumber\\
\label{L1eq}
&&L_1(z)=8(1-kz)z^2p''(z)-2(3+kz)zp'(z)+(3-kz)p(z)\\
&&\qquad\qquad
+2(3-kz)zq'(z)-(9-kz)q(z)\nonumber\\
&&\qquad\qquad
+16(1-kz)z^3h_1''(z)+16(1-2kz)z^2h_1'(z)-4(1+kz)zh_1(z)\\
&&\qquad\qquad
+8(1-kz)z^3h_0''(z)+4(2-5kz)z^2h_0'(z)-2(1+2kz)zh_0(z)\nonumber\\
\label{L2eq}
&&L_2(z)=-4(2-kz)zp'(z)+2(1-kz)p(z)-6q(z)
\nonumber\\
&&\qquad\qquad-6(3-kz)z^2h_1'(z)-3(3-kz)zh_1(z)
-6(1-kz)z^2h_0'(z)-3(1-kz)zh_0(z)\\
\label{L3eq}
&&L_3(z)=32(1-kz)z^2p''(z)-32zp'(z)+2(25-4kz)p(z)\nonumber\\
&&\qquad\qquad+20(2-kz)zq'(z)-10(7-kz)q(z)\nonumber\\
&&\qquad\qquad
+120(1-kz)z^3h_1''(z)+30(3-7kz)z^2h_1'(z)-15(3+kz)zh_1(z)\\
&&\qquad\qquad+40(1-kz)z^3h_0''(z)+30(1-3kz)z^2h_0'(z)-15(1+kz)zh_0(z)\nonumber\\
\label{L4eq}
&&L_4(z)=-8(1-kz)z^2w''(z)+8zw'(z)+2kzw(z).
\end{eqnarray}

The equations in \eref{Leqs} and \eref{L0eq}--\eref{L1eq} look quite 
complicated but it turns out that they can be further simplified. First
we note that $w(z)$ has decoupled from the other functions and
only appears in $L_4$.  To solve for this, write $w(z)=z^{1/2}\widetilde w(z)$, where
$\widetilde w(z)$ is the function in the inertial frame.  The last equation in
\eref{Leqs} can then be rewritten as
\begin{equation}\label{weq}
\widetilde w''(z)-\frac{k}{1-kz}\widetilde w'(z)-\frac{3}{4z^2}\widetilde w(z)=
-\frac{A(z)-B(z)}{8z^{1/2}(1-kz)}.
\end{equation}
Using \eref{ABeq} we see that  this is the same equation as in 
\eref{inhom}, except the source term in \eref{weq} has an extra factor 
of $-24$.  Hence the solution for $\widetilde w(r)$ is
\begin{equation}\label{wsol}
\widetilde w(r)=-24\phi(r),
\end{equation}
where $\phi(r)$ is given in \eref{phisol}.  The expression in \eref{phisol}
 is  quite
complicated, but as we will see, this does not matter since
$w(r)$  does not contribute to the lowest order correction to the entropy. 

Given \eref{weq} and the arguments in the previous section, 
it  follows that the homogeneous mode for $\widetilde w(r)$ 
corresponds to a dimension 6 operator.  Consulting
Table III of \cite{krvn}, we see that this is the scalar in the
{\bf 84} representation of $SO(6)$ with $M^2=12$.
The identity of the two fluctuation equations \eref{inhom} and \eref{weq},
including their
sources, suggests that the two fields are in the same supermultiplet, 
although it is not clear why the sources should agree
for $T$ non-zero.

Let us next consider the linear combination
\begin{equation}
\frac{5}{8}L_0(z)+\frac{15}{8}L_1(z)+\frac{5}{8}L_2(z)-\frac{3}{8}L_3(z)
=
\frac{19}{24}A(z)-\frac{3}{8}B(z).
\end{equation}
This reduces to the equation
\begin{equation}\label{F1eq}
8(1-kz)z^2p''(z)-8zp'(z)-2(5+kz)p(z)=
\frac{19}{24}A(z)-\frac{3}{8}B(z).
\end{equation}
The solution to this equation with the correct boundary conditions is 
surprisingly simple,
\begin{equation}\label{psol}
p(z)=-\frac{1}{24}A(z).
\end{equation}
If $T=0$, then the homogeneous solutions for $z^{-1/2}p(z)$
 behave like $z^{2}$ and
$z^{-1}$.  Hence this mode corresponds to a dimension 8 operator.  Again
consulting Table III of \cite{krvn}, this is the $SO(6)$ 
singlet from ${h^a}_a$ with $M^2=32$.

Next, we find that
the first two equations and the derivative of the third equation in \eref{Leqs}
are linearly
dependent and that consistent solutions exist provided that the following
equation is satisfied:
\begin{equation}\label{check}
(1-kz)\left[2zA'(z)+6zB'(z)-A(z)\right]-3(5-3kz)B(z)=0.
\end{equation}
One can easily verify that the expressions for $A(z)$ and $B(z)$ in \eref{ABeq}
satisfy \eref{check}.  

Of course \eref{check} must be true by general covariance;  the solutions for
$A(z)$ and $B(z)$ came from the equations of motion for $G_{\it3}$.  These
were derived from the same generally covariant action that the Einstein
equations were derived from.
The fact that there are only four linearly independent equations means that
there is an extra gauge degree of freedom.  Thus,  we can
choose one of the unknown functions to be an arbitrary function.  
A particularly
useful gauge choice is 
\begin{equation}\label{gauge}
h_1(z)=h_0(z).
\end{equation}
Using \eref{gauge}, the third equation in \eref{Leqs} reduces to
\begin{equation}\label{h0f21}
-(2-kz)\left(12z^2h_0'(z)+6zh_0\right)-6q(z)=(2-kz)\left(-\frac{1}{6}zA'(z)
+\frac{1}{12}A(z)\right)+\frac{1}{4}B(z),
\end{equation}
where we have substituted the solution in \eref{psol}.  Substituting 
\eref{gauge} into the difference between the 
second and first equations in \eref{Leqs} gives
\begin{equation}\label{h0f22}
k\left(16z^3 h_0'(z)+ 8z^2h_0(z)-4z^2q'(z)+2zq(z)\right)=
k\left(\frac{1}{6}z^2A'(z)-\frac{1}{12}zA(z)\right).
\end{equation}
Examining \eref{h0f21} and \eref{h0f22}, we see that the same linear 
combination of $h_0(z)$ and $h_0'(z)$ appears, so we can reduce these
equations to a single first order equation for $q(z)$.  Using \eref{check}
we can replace all $A(z)$ dependent terms by $B(z)$ and $B'(z)$.  
The equation can be solved using standard methods, giving the solution
\begin{equation}\label{qsol}
q(z)=\frac{1}{24}\left(-B(z) +\frac{2-kz}{\sqrt{z}}
\int_0^z\frac{\s(z')}{\sqrt{z'}}dz'\right),
\end{equation}
where $\s(z)$ is defined in \eref{ABeq}.  An integration constant has
been chosen so that $q(z)$ has the correct behavior on the boundary.

To solve for $h_0(z)$ one just substitutes the solution for $q(z)$ in 
\eref{qsol} into \eref{h0f21}.  Using \eref{check} and integrating by parts,
one finds
\begin{equation}\label{h0sol}
h_0(z)=\frac{1}{96z}\left(A(z)-B(z) +\frac{2-kz}{\sqrt{z}}
\int_0^z\frac{\s(z')}{\sqrt{z'}}dz'\right).
\end{equation}
Clearly, \eref{psol}, \eref{qsol} and \eref{h0sol} satisfy
\begin{equation}\label{solrel}
p(z)+4zh_0(z)-q(z)=0.
\end{equation}

Using these solutions, we can go back and find the corrections to $\chi_{\it4}$,
the space-time components of $C_{\it4}$.
 From \eref{G3eq3} and \eref{chi1eq}, we find that the change in $\chi_{\it4}$ is
\begin{eqnarray}\label{C4eq}
\de \chi_{\it4}\*\frac{1}{2g}\int^z\frac{dz}{z^{5/2}}\left(\frac{1}{12}A(z)+
[p(z)-q(z)-4zh_0(z)]\right)dx^0\wedge dx^1\wedge dx^2\wedge dx^3\nonumber\\
\*\frac{1}{24g}\int^z\frac{dz'}{{z'}^{5/2}}\left(B(z')
-\frac{2-kz'}{\sqrt{z'}}
\int_0^{z'}\frac{\s(z'')}{\sqrt{z''}}dz''\right)
dx^0\wedge dx^1\wedge dx^2\wedge dx^3.
\end{eqnarray}
The last term can be integrated by parts.  Using the relation in \eref{ABeq},
one is left with
\begin{equation}\label{C4sol}
\de \chi_{\it4}=\frac{1-kz}{24gz^2}\int_0^z\frac{\s(z')}{\sqrt{z'}}dz'
dx^0\wedge dx^1\wedge dx^2\wedge dx^3.
\end{equation}
In particular, we note that $\de \chi_{\it4}=0$ on the horizon, verifying the claim
in section 2 that the $r_0^2$ term in the probe action goes to zero at the
horizon.  

A useful check of this analysis is to explicitly compute the $m^2$ terms that
appear in \eref{d5pot} at zero temperature.  This term comes from  corrections 
to $C_{\it4}$ and the parallel components of the metric.  
Inserting these corrections into \eref{D5action}, one finds that the correction
to $-S/V$ is
\begin{equation}\label{m2SV}
4\pi^2n\mu_5\a'\left[\frac{2}{g\sqrt{z}}h_0(z)-\de C_{0123}(z)\right].
\end{equation}
Using equations \eref{h0sol} and \eref{C4sol}, the expression in
\eref{m2SV} reduces to
\begin{equation}\label{m2SV2}
\frac{\pi^2n\mu_5\a'}{12g}\left(
A(z)-B(z)+\frac{k}{z}\int_0^z\frac{\s(z')}{\sqrt{z'}}dz'\right)
\end{equation}
In the zero temperature limit we set $k=0$.
Using \eref{ABeq} and the relation
\begin{equation}\label{T3m2}
|T_{\it 3}|^2=18m^2,
\end{equation}
one finds that the correction in \eref{m2SV2} is
\begin{equation}\label{m2SV3}
\frac{2\pi^2n\mu_5\a'}{g}m^2r^2,
\end{equation}
which is precisely the term required by supersymmetry.	

We close this section by listing the inhomogeneous solutions for the $T=0$
case.  Setting $k=0$ in \eref{phisol}, \eref{wsol},
\eref{psol}, \eref{qsol} and 
\eref{h0sol}, we find for the gauge choice \eref{gauge}
\begin{eqnarray}\label{inhomsolT0}
w(r)\*-\frac{3m^2R^6}{r^4}\nonumber\\
p(r)\*-\frac{9m^2R^6}{8r^4}\nonumber\\
q(r)\*+\frac{m^2R^6}{24r^4}\\
h_0(r)\*+\frac{7m^2R^2}{24}\nonumber.
\end{eqnarray}

\section{Entropy Corrections}

In the low temperature phase the masses of the hypermultiplets are much larger
than the temperature and therefore the contribution of the hypermultiplets
to the entropy is exponentially suppressed.  In the high temperature phase,
where $T>>m$, the masses give polynomial corrections to the entropy.

At weak coupling, the entropy density $\cal S$ is given by
\begin{equation}\label{entropyweak}
{\cal S}=\frac{2\pi^2}{3} N^2T^3-\frac{3}{4}N^2m^2T+{\rm O}(m^4).
\end{equation}
In \cite{gkp96} it was shown that at strong coupling, 
the leading order term, ${\cal S}_0$, is 
\begin{equation}\label{S0eq}
{\cal S}_0=\frac{\pi^2}{2}N^2T^3,
\end{equation}
which is $3/4$ the value of the leading term
in \eref{entropyweak}.  In this section we will compute the next term in
the series at strong coupling. 

To compute the entropy correction, one needs to compute the ${\rm O}(m^2)$ 
correction
to the horizon area and the temperature for a fixed Schwarzschild
radius. Using the metric in \eref{metcorr}, we see that 
the ${\rm O}(m^2)$ correction to the area $\cal A$ for a fixed $r_H$ is
\begin{equation}\label{areacorr}
\Delta {\cal A} =\frac{1}{2}g^{\hat\mu\hat\nu}\Delta g_{\hat\mu\hat\nu}
{\cal A}_0=\frac{1}{2}\frac{r_H^2}{R^2}\left(3h_1(r_H)\frac{R^4}{r_H^4}
+p(r_H)\right){\cal A}_0,
\end{equation}
where the $\hat\mu$ indices refer to the coordinates orthogonal to the temporal
and radial directions.  
${\cal A}_0$ is the usual $AdS_5\times S_5$ Schwarzschild area.

Because of the corrections to the metric, it is necessary to adjust the 
circumference of the Euclidean time circle in order to prevent a conical 
singularity at $r=r_H$.
The corresponding correction to the temperature, for fixed $r_H$, is 
\begin{equation}\label{tempcorr}
\Delta T = \frac{1}{2}\frac{r_H^2}{R^2}\left(h_0(r_H)\frac{R^4}{r_H^4}-q(r_H)
\right)T.
\end{equation}
Hence, using \eref{solrel} and \eref{S0eq}, we find that the ${\rm O}(m^2)$
correction to the entropy for fixed {\it temperature} is
\begin{eqnarray}\label{entropycorr}
\Delta {\cal S}\* \left(\frac{\Delta {\cal A}}{{\cal A}_0}
-3\frac{\Delta T}{T}\right){\cal S}_0\nonumber\\
\* 2\frac{r_H^2}{R^2}\left(3h_0(r_H)\frac{R^4}{r_H^4}
+p(r_H)\right){\cal S}_0.
\end{eqnarray}

If we now use the relations in \eref{ABeq}, \eref{psol}, \eref{h0sol} and
\eref{T3m2},
we find that
\begin{equation}\label{entropycorr2}
\Delta {\cal S} = -\frac{9}{16}\frac{\Gamma(3/4)^4}{\pi^3}\frac{m^2}{T^2}
\left(1-
\frac{2}{3}\int_0^1\frac{x^2\left[P_{-1/2}(x)-xP_{1/2}(x)\right]^2}{(1-x^2)^2}
dx\right){\cal S}_0.
\end{equation}
We are unaware of any analytic expression for the above integral, but the
numerical result for $\Delta {\cal S}$ is
\begin{equation}\label{DelS}
\Delta {\cal S} = - 0.1714 N^2 m^2T,
\end{equation}
which is 23\% 
of the weak coupling result.  

This demonstrates the shrinking of the horizon area as the mass is increased.
It also suggests that a critical temperature is reached when $T\sim m$.
It is hoped that one can carry out a Hawking-Page analysis of this
 transition \cite{Hawking:1983dh,Witten:1998zw}, by demonstrating that the 
free energy is lowered if
the black-hole geometry of the bulk
changes  to the Polchinski-Strassler geometry. However, since the full 10 
dimensional space
is not a product space, in order to proceed one will
require a more sophisticated approach for evaluating
 boundary terms in the Euclidean action.  

As was case for the $T^3$ term, the $m^2T$ term in the entropy is 
somewhat suppressed at strong coupling.  In fact as a percentage, the 
suppression is more.  An interesting test would be to carry out the
analysis in \cite{gkt} for the \noneplus\ system.  In \cite{gkt}, the first
order correction in $1/(gN)$ was computed, and it was shown that the entropy
increases when moving away from the strong coupling limit.  For the \noneplus\ 
system, one should find that the absolute value of the $m^2T$ term also 
increases when moving away from strong coupling.

\section{Discussion}

The results presented for the backreaction required a lot of tedious 
calculation.  But in the end, the results had significant simplification.  
This leaves us optimistic that an exact supergravity solution can be found.
Recent progress in finding exact solutions for other supergravity
duals \cite{Pilch:2000ej,Pilch:2000ue,Brandhuber:2000ct,Pilch:2000fu,Klebanov:2000rd,Klebanov:2000nc,Oh:2000ez,Klebanov:2000hb} should give one hope in 
succeeding in this case as well.

It might be the case that finding an exact solution for the high temperature 
phase will be an easier task then finding the solution for the
the low temperature phase.  While the low temperature phase has
unbroken \none\ supersymmetry , it does have the extra complication of
5 branes wrapping 2-spheres.   In any case, armed with an exact solution
of the high temperature phase, one should be able to observe the shrinking
of the black hole horizon down to zero area for some nonzero $T=T_c$.
Work on this problem continues.

\acknowledgements{}
We thank L. Rastelli, M. Strassler and N. Warner for  helpful 
discussions.
This research was supported in part by the NSF under grant number
PHY-97-22072 and by the U.S. Department of Energy under contract number
DE-FC02-94ER40818.

\section{Appendix}

To get a better estimate for the temperature at which a metastable state
disappears, we need to compute a discriminant.  The derivative of
\eref{d5pot} is proportional to 
\begin{equation}
4Ar^3\left(\sqrt{1-r_H^4/r^4}-1\right)\ +\ \frac{2Ar_H^4}{r\sqrt{1-r_H^4/r^4}}
\ +\ 2(r^2-rr_0)(2r-r_0),
\end{equation}
where $A=(\pi n^2)/(2gN)>>1$.  Hence, to find extrema for the potential
one looks for zeros of the polynomial
\begin{eqnarray}
(8A-4)r^8+(12-12A)r^7+(-13+4A)r^6+6r^5+(-1-8Ar_H^4+4r_H^4)r^4\nonumber\\+
(-12r_H^4+12Ar_H^4)r^3
+(13r_H^4-4Ar_H^4)r^2-6r_H^4r+r_H^4+A^2r_H^8.
\end{eqnarray}
In fact, we want to find where two extrema coalesce, which occurs when the
discriminant of this polynomial has a zero.  The discriminant factors into
a seventh order polynomial and and a second order polynomial in $r_H^4$, and
it turns out that it is the zeros of the seventh order polynomial that we need.
However, for large $A$ it turns out that this polynomial is approximately
\begin{equation}
2^{29}A^9(r_H^4)^7-3^3(3351797)A^8r_0^8(r_H^4)^5-2^83^6r_0^{16}(r_H^4)^3.
\end{equation}
Hence, there is a zero at
\begin{equation}
r_H\approx \frac{1}{16}3^{3/8}2^{1/4}\left(3351797+(171)(2659)\sqrt{57}\right)^{1/8} A^{-1/8}r_0,
\end{equation}
and so the solution disappears at
\begin{equation}
r_H\approx \kappa\left(\frac{n^2}{4\pi gN}\right)^{-1/8}r_0,
\end{equation}
where
\begin{equation}
\kappa=\frac{1}{16}\left(\frac{2}{\pi^2}\right)^{1/8}3^{3/8}
\left(3351797+(171)(2659)\sqrt{57}\right)^{1/8}\approx 0.5522
\end{equation}
In terms of the temperature and $m$, this is 
\begin{equation}
T=\kappa\left(\frac{n^2}{4\pi gN}\right)^{3/8}|m|.
\end{equation}

\end{document}